\documentclass[12pt]{article}

\usepackage{amssymb}
\usepackage{amsmath}
\usepackage{latexsym}
\usepackage{yfonts}

\catcode`@=11
\renewcommand{\section}{\@startsection{section}{1}{0pt}{\medskipamount}
{\medskipamount}{\large\bf}} \numberwithin{equation}{section}
\catcode`@=12
\def\beq{\begin{eqnarray}}    %%%  begequation/eqnarray
\def\eeq{\end{eqnarray}}      %%%  endequation/eqnarray

\def\={\ =\ }

\begin{document}

\begin{center}

{\Large\bf BRST-BV quantization of gauge theories\\ with global
symmetries}

\vspace{18mm}

{\large I. L. Buchbinder $^{(a, b, c)}\footnote{E-mail:
joseph@tspu.edu.ru}$, P. M. Lavrov$^{(a, b)}\footnote{E-mail:
lavrov@tspu.edu.ru}$\; }

\vspace{8mm}

\noindent  ${{}^{(a)}} ${\em
Center of Theoretical Physics, \\Tomsk State Pedagogical University,\\
Kievskaya St.\ 60, 634061 Tomsk, Russia}

\noindent  ${{}^{(b)}} ${\em
National Research Tomsk State  University,\\
Lenin Av.\ 36, 634050 Tomsk, Russia}

\noindent  ${{}^{(c)}} ${\em
Departamento de F\'isica, ICE,\\ Universidade Federal de Juiz de Fora,\\
Campus Universit\'ario-Juiz de Fora,\\ 36036-900, MG, Brazil}

\vspace{20mm}

\begin{abstract}
\noindent We consider the general gauge theory with a closed
irreducible gauge algebra possessing the non-anomalous global
(super)symmetry in the case when the gauge fixing procedure violates
the global invariance of classical action. The theory is quantized
in the framework of BRST-BV approach in the form of functional
integral over all fields of the configuration space. It is shown
that the global symmetry transformations are deformed in the process
of quantization and the full quantum action is invariant under such
deformed global transformations in the configuration space. The
deformed global transformations are calculated in an explicit form
in the one-loop approximation.

\end{abstract}

\end{center}

\vfill

\\

\newpage

\section{Introduction}

BRST quantization procedure, initiated in the works
\cite{BRS,T,FvNF,DZ,Nie1,Kal,Town,dWvH}, which can be realized
within the Hamiltonian BFV approach \cite{FV,BV2} or within the
Lagrangian BV approach \cite{BV}, \cite{BV1}, is a powerful and
universal tool to formulate the quantum gauge theories and
investigate their structure. This procedure is applicable to an
extremely wide class of gauge theories including the
(super)Yang-Mills theories, (super)gravity, (super)strings and more
specific gauge theories like the gauge antisymmetric field models.
All these theories possess many common properties therefore one can
talk about a general gauge theory and study its quantum aspects in
general terms.

Besides the local symmetries, the many gauge theories are
characterized by the rigid symmetries. For example,  the role which
is played by the global chiral symmetry in the Standard Model (see
e.g. \cite{standard model}) or the role which is played by global
conformal group in the String Theory (see e.g. \cite{string theory})
are well known. In many cases it is essential to preserve the
classical global symmetry in quantum theory since in the opposite
case the important physical properties of the theory can be
violated. Example of such a model is ${\cal N}=4$ supersymmetric
Yang-Mills theory, where the classical rigid superconformal symmetry
is conserved at quantum level (see e.g. \cite{N=4}). As a result, we
arrive at a general problem to study the classical global symmetries
at BRST quantization of the gauge theory.

A specific aspect of the above general problem arises in
supersymmetric gauge theories. These theories possess  a gauge
symmetry and rigid or global supersymmetries. Such theories can be
formulated either in the component formalism or in the terms of
superfields. In the first case the supersymmetry is non manifest and
when we quantize the corresponding gauge theory imposing the gauge
fixing conditions on vector fields we violate the rigid
supersymmetry. Therefore it is unclear if the quantum effective
action should be supersymmetric. At first, this problem arose in
${\cal N}=1$ super Yang-Mills theories and led to the study of the
aspects of global symmetry in quantum gauge theories
\cite{dWF,BPT,vanH,BHW} However, this problem was automatically
resolved after formulation of the ${\cal N}=1$ supergauge theories
in terms of ${\cal N}=1$ superfields (see e.g. \cite{BK}), where the
manifestly ${\cal N}=1$ gauges have been used. However, the problem
is still remains in extended super Yang-Mills theories. At present,
the best formulation of $4D, {\cal N}=4$ rigid super Yang-Mills
theory is achieved in terms of $4D, {\cal N}=2$ or in terms of $4D,
{\cal N}=3$ harmonic superfields \cite{GIOS}. In this case only part
of supersymmetries are manifest, the other remain hidden. The same
situation will be in $6D, {\cal N}=(1,1)$ rigid super Yang-Mills
theory, which is similar in many aspects to $4D, {\cal N}=4$ super
Yang-Mills theory. At present, the best formulation of such a theory
is achieved in terms of $6D, {\cal N}= (1,0)$ superfields (see e.g.
\cite{BIS} and references therein). Again, one part of
supersymmetries is manifest and the other part is hidden. In all
such theories the gauges preserve the manifest symmetries and
violate the hidden supersymmetries (see
e.g.\cite{BK1,BK2,BIP,BIS,BIMS,BIMS1}).

Another aspect of the same problem arises at quantization of $4D,
{\cal N}=2$ superconformal theories where the gauges preserve the
manifest ${\cal N}=2$ supersymmetry but violate the global
superconformal invariance. This aspect was studied in the series of
the papers \cite{K1,K2,K3} where an approach to general problem of
global symmetries in quantum gauge theories was proposed and it was
shown under some assumptions concerning the structure of the theory
that the quantization leads to deformation of global symmetries and,
in particular, to deformation of superconformal transformations.

In this paper we study the above problem in general terms and
generalize all the previous results. We consider the general gauge
theory possessing the non-anomalous global symmetry and quantize
this theory in the framework of the BRST-BV procedure. It is assumed
that gauges violate the initial global symmetry of the classical
action. Under these conditions we develop maximally general approach
to structure of the global symmetries at the quantization of gauge
theories. We show like in work \cite{K1} that the quantization
procedure leads to deformation of classical global transformations.
However, we use the more general effective action then in \cite{K1}
what provides more possibilities for purely algebraic analysis of
the global symmetries in quantum gauge theory. We prove that
although the vacuum functional is invariant under classical global
transformations, the effective action is not invariant and its
invariance requires the quantum corrections to generators of global
symmetry transformations. The maximally general form of such
deformations is derived. In particular, we prove that the full
quantum action in the functional integral is invariant under the
deformed transformations. The one-loop invariance is studied in
details and the corresponding one-loop deformation of the global
transformations is calculated in explicit form. The results obtained
are exclusive general and fulfill for any bosonic or fermionic gauge
theory with an irreducible closed gauge algebra and with an
arbitrary (even open) non-anomalous algebra of global symmetry.

The paper is organized as follows. Section 2 is devoted to
description of the general gauge theory with closed irreducible
gauge algebra and fixing the notations and conventions. In Section 3
we describe the properties of the general gauge theory with global
symmetry. In Section 4 we construct the quantum action of the theory
under consideration and prove that the quantization leads to
deformation of the classical global symmetries. In section 5 we
consider the above deformation in one-loop approximation and
construct the deformation in an explicit form. Section 6 summarizes the
results.

In the paper the DeWitt's condensed notations are used \cite{DeWitt}.
We employ the notation $\varepsilon(A)$ for the
Grassmann parity of any quantity $A$. All derivatives with respect
to sources are taken from the left only.  The right and left
derivatives with respect to fields are marked by special
symbols $"\leftarrow"$  and $"\rightarrow"$ respectively. The symbol
$A_{,i}(\varphi)$ means the right derivative of $A(\varphi)$
with respect to the field $\varphi^i$.

\setcounter{section}{1}
\renewcommand{\theequation}{\thesection.\arabic{equation}}
\setcounter{equation}{0}

\section{General gauge theory: notations and conventions}

Consider the general gauge theory with closed irreducible gauge
algebra. It means that the  initial action, $S_0=S_0(\varphi)$, of the
fields $\varphi=\{\varphi^i\},\;
i=1,2,...,n,\;\varepsilon(\varphi^i)=\varepsilon_i$ is invariant
under the gauge transformations \beq \label{g1}
S_{0,i}(\varphi)R^i_{\alpha}(\varphi)=0,\quad \delta
\varphi^i=R^i_{\alpha}(\varphi)\xi^{\alpha}, \eeq where
$\xi^{\alpha}$ are the arbitrary functions with Grassmann parities
$\varepsilon(\xi^{\alpha})\equiv\varepsilon_{\alpha}$, and
$R^i_{\alpha}=R^i_{\alpha}(\varphi)$,
$\varepsilon(R^i_{\alpha})=\varepsilon_i + \varepsilon_{\alpha}$ are
generators of gauge transformations which are assumed to be linear
independent in gauge indices $\alpha$. It is convenient to introduce
the operators of gauge transformations, ${\hat R}_{\alpha}={\hat
R}_{\alpha}(\varphi)$, \beq \label{g1a} {\hat
R}_{\alpha}=\frac{\overleftarrow{\delta}}{\delta
\varphi^i}R^i_{\alpha}, \eeq so that the gauge invariance of $S_0$
(\ref{g1}) is written in the form \beq \label{g1b} S_0{\hat
R}_{\alpha}=0. \eeq

The algebra of the gauge generators ${\hat R}_{\alpha}$ has the
following form due to the closure:
\beq
\label{g1c}
[{\hat R}_{\alpha}, {\hat R}_{\beta}]=-
{\hat R}_{\gamma}{F^{\gamma}}_{\!\!\alpha\beta}
\eeq
or
\begin{eqnarray}
\label{g2} R^i_{\alpha ,
j}R^j_{\beta}-(-1)^{\varepsilon_{\alpha}\varepsilon_
{\beta}}R^i_{\beta ,j}R^j_{\alpha}
=-R^i_{\gamma}{F^{\gamma}}_{\!\!\alpha\beta},
\end{eqnarray}
where ${F^{\gamma}}_{\!\!\alpha\beta}={F^{\gamma}}_{\!\!\alpha\beta}(\varphi)$
are the structure coefficients  depending
in general on fields $\varphi$ and obeying the symmetry properties
\beq
\label{g2aa}
{F^{\gamma}}_{\!\!\alpha\beta}
=-(-1)^{\varepsilon_{\alpha}\varepsilon_{\beta}}
{F^{\gamma}}_{\!\!\beta\alpha}.
\eeq
For Yang-Mills  theories they are constants.
The Jacobi identities written in terms of the gauge generators
and the structure coefficients read
\beq
\label{g2a}
\big({F^{\sigma}}_{\!\!\alpha\rho}{F^{\rho}}_{\!\!\beta\gamma}+
{F^{\sigma}}_{\!\!\alpha\beta,i}R^i_{\gamma}\big)
(-1)^{\varepsilon_{\alpha}\varepsilon_{\gamma}}
+{\rm cycle (\alpha\beta\gamma)}=0.
\eeq

For irreducible gauge theories
the extended configuration space is described by the fields :
\beq
\label{g3}
&&\phi^A=(\varphi^i,\;B^{\alpha},\;C^{\alpha},\;\bar{C}^{\alpha}),
\\
&&\varepsilon(\varphi^i)=\varepsilon_i,\;\;
\varepsilon(B^{\alpha})=\varepsilon_{\alpha},\;\;
\varepsilon(C^{\alpha})=
\varepsilon(\bar{C}^{\alpha})=\varepsilon_{\alpha}+1,
\nonumber\\
&&{\rm gh}(\varphi^i)={\rm gh}(B^{\alpha})=0,\quad {\rm gh}(C^{\alpha})=1, \quad
{\rm gh}\bar{C}^{\alpha})=-1,
\nonumber
\eeq
where $B^{\alpha}$ are Nakanishi-Lautrup auxiliary fields,
$C^{\alpha}$ and $\bar{C}^{\alpha}$ are the ghost and anti-ghost fields.
For gauge theories under consideration which belong to the rank 1
gauge theories in terminology of BV formalism \cite{BV,BV1}
the total (quantum) action, $S=S(\phi)$,
can be written in the form of the Faddeev-Popov action \cite{FP},
\beq
\label{g4}
S(\phi)=S_0(\varphi)+\bar{C}^{\alpha}\chi_{\alpha
,i}(\varphi)R^i_{\beta}(\varphi)C^{\beta} +\chi_{\alpha}(\varphi)B^{\alpha}
\eeq
where $\chi_{\alpha}=\chi_{\alpha}(\varphi),\;\varepsilon(\chi_{\alpha})=\varepsilon_{\alpha},$
are some gauge functions lifting the degeneracy of the
classical gauge invariant action $S_0$.

The action (\ref{g4}) is invariant under the following BRST
transformation
\beq
\label{g5}
 \delta_{\rm B} S(\phi)=0,\quad \delta_{\rm B}\phi^A=R^A(\phi)\lambda
\eeq
with
\beq
\label{g6}
R^A(\phi)=(R^i_{\alpha}(\varphi)C^{\alpha},\;0\;,
-\hbox{\large$\frac{1}{2}$}(-1)^{\varepsilon_{\beta}}
{F^{\alpha}}_{\!\!\beta\gamma}(\varphi)C^{\gamma}C^{\beta}, B^{\alpha}),
\eeq
where $\lambda$ is a constant Grassmann parameter
($\varepsilon(\lambda)=1$).
Introducing the operator of BRST transformations,
${\hat R}={\hat R}(\phi)$, and using abbreviation $R^A=R^A(\phi)$
\beq
\label{g6a}
{\hat R}=\frac{\overleftarrow{\delta}}{\delta \phi^A}R^A,\quad
\varepsilon({\hat R}) =1,
\eeq
the BRST invariance of action $S$ (\ref{g5}) can be written as
\beq
\label{g6b}
S{\hat R}=0.
\eeq
With the help of ${\hat R}$ the action (\ref{g4}) rewrites in the form
\beq
\label{g6c}
S=S_0+\Psi{\hat R},
\eeq
where we have introduced the gauge fixing functional $\Psi=\Psi(\phi)$ of the form
\beq
\label{g6d}
\Psi(\phi)={\bar C}^{\alpha}\chi_{\alpha}(\varphi),\quad
\varepsilon(\Psi)=1.
\eeq
The quantum action in form of (\ref{g6c}) is evidently BRST invariant
due to the nilpotency of ${\hat R}$, ${\hat R}^2=0$.

\section{General gauge theory with rigid symmetry}

%Now let us suppose the invariance of initial action $S_0(\varphi)$
%under a global symmetry transformations as well,
Let the initial action $S_0(\varphi)$ is invariant
under  the global symmetry transformations as well,
\beq \label{g7}
S_{0,i}(\varphi)T^i_{a}(\varphi)=0,\quad \delta_T
\varphi^i=T^i_a(\varphi)\omega^a \eeq where $\omega^a, \;
a=1.2,...,m$ are constant parameters  with Grassmann parities
$\varepsilon(\omega^a)\equiv\varepsilon_a$, and
$T^i_a=T^i_a(\varphi)$ ($\varepsilon(T^i_a)=\varepsilon_i +
\varepsilon_a$) are generators of global transformations.  Let
${\hat T}={\hat T}(\varphi)$, \beq \label{g7a} {\hat T}={\hat
T}_a\omega^a,\quad {\hat T}_a={\hat
T}_a(\varphi)=\frac{\overleftarrow{\delta}}{\delta
\varphi^i}T^i_{a}, \eeq be the operator of global transformations.
Then, in general (see \cite{BHW}) \beq \label{g7b} [{\hat T}_a,{\hat
T}_b]=-{\hat T}_c{f^{c}}_{\!\!ab}- {\hat
R}_{\alpha}{K^{\alpha}}_{\!\!ab}-\frac{\overleftarrow{\delta}}{\delta\varphi^i}
\lambda^{ij}_{\;\;ab} S_{0,j}, \eeq or \beq \label{g7c} T^i_{a ,
j}T^j_{b}-(-1)^{\varepsilon_a\varepsilon_b}T^i_{b ,j}T^j_{a}
=-T^i_{c}{f^{c}}_{\!\!a b}-R_{\alpha}^i{K^{\alpha}}_{\!\!ab}-
\lambda^{ij}_{\;\;ab} S_{0,j}, \eeq Here, \beq \label{h1}
&&{f^{c}}_{\!\!ab}={f^{c}}_{\!\!ab}(\varphi),\quad {f^{c}}_{\!\!ab}=
-{f^{c}}_{\!\!ba}(-1)^{\varepsilon_a\varepsilon_b}, \quad
\varepsilon({f^{c}}_{\!\!ab})=
\varepsilon_a+\varepsilon_b+\varepsilon_c,\\
\label{h2}
&&{K^{\alpha}}_{\!\!ab}={K^{\alpha}}_{\!\!ab}(\varphi),\quad
{K^{\alpha}}_{\!\!ab}=-{K^{\alpha}}_{\!\!ba}(-1)^{\varepsilon_{a}\varepsilon_b},
\varepsilon({K^{\alpha}}_{\!\!ab})=\varepsilon_{a}+\varepsilon_b+\varepsilon_{\alpha},\\
\label{h3}
&&\lambda^{ij}_{\;\;ab}=\lambda^{ij}_{\;\;ab}(\varphi),\quad
\lambda^{ij}_{\;\;ab}=-\lambda^{ij}_{\;\;ba}(-1)^{\varepsilon_a\varepsilon_b},\quad
\varepsilon(\lambda^{ij}_{\;\;ab})=\varepsilon_a+\varepsilon_b+\varepsilon_i+\varepsilon_j.
\eeq

To close the algebra of gauge and global symmetries (\ref{g2}) and (\ref{g7c})
we  add the following relations
\beq
\label{g7d1}
[{\hat R}_{\alpha},{\hat T}_{a }]=-{\hat R}_{\beta}{U^{\beta}}_{\!\!\alpha a},
\eeq
or
\beq
\label{g7e}
R^i_{\alpha, j}T^j_{b}-(-1)^{\varepsilon_{\alpha} \varepsilon_b}T^i_{b ,j}R^j_{\alpha}
=-R^i_{\beta}{U^{\beta}}_{\!\!\alpha b},
\eeq
which means that the commutator  of gauge and global transformations is characterized
by local parameters.
The structure coefficients
${U^{\beta}}_{\!\!\alpha b}={U^{\beta}}_{\!\!\alpha b}(\varphi)$ obey the symmetry properties
\beq
\label{g7f}
{U^{\beta}}_{\!\!\alpha b}=
-{U^{\beta}}_{\!\!b \alpha}(-1)^{\varepsilon_{\alpha}\varepsilon_b}.
\eeq

The Jacobi identities for global generators have the form
\beq
\nonumber
&&T^i_d\big({f^{d}}_{\!\!ae}{f^{e}}_{\!\!bc}+{f^{d}}_{\!\!a b,i}T^i_{c}\big)
(-1)^{\varepsilon_{a}\varepsilon_c}+
R^i_{\alpha}\big({U^{\alpha}}_{\!\!a\beta}
{K^{\beta}}_{\!\!bc}+{K^{\alpha}}_{\!\!a b,i}T^i_{c}\big)
(-1)^{\varepsilon_{a}\varepsilon_c}
+\\
\nonumber
&&+\big(T^i_{a,j}\lambda^{j\;\!k}_{\;\;bc}S_{0,k}-
\lambda^{ij}_{\;\;ab}S_{0,jk}T^k_c-
\lambda^{ij}_{\;\;ab,k} S_{0,j}
T^k_c(-1)^{\varepsilon_j\varepsilon_k}\big)(-1)^{\varepsilon_{a}\varepsilon_c}+\\
\label{g7d0} &&+{\rm cycle} \;(a, b, c) =0. \eeq In the case of
closed global transformations ($\lambda^{ij}_{\;\;ab}=0$) the Jacobi
identity (\ref{g7d0}) splits into two relations \beq
&&\big({f^{d}}_{\!\!ae}{f^{e}}_{\!\!bc}+{f^{d}}_{\!\!a
b,i}T^i_{c}\big)
(-1)^{\varepsilon_{a}\varepsilon_c}+{\rm cycle} \;(a, b, c) =0,\\
&&\big({U^{\alpha}}_{\!\!a\beta}
{K^{\beta}}_{\!\!bc}+{K^{\alpha}}_{\!\!a b,i}T^i_{c}\big)
(-1)^{\varepsilon_{a}\varepsilon_c}+{\rm cycle} \;(a, b, c) =0.
\eeq

Using the Jacobi identity for two global and one gauge
transformations we obtain the following relations
\beq
\nonumber
&&T^i_c{f^{c}}_{\!\!a b,i}R^i_{\alpha}(-1)^{\varepsilon_{\alpha}\varepsilon_a}+
R^i_{\gamma}\big(({U^{\gamma}}_{\!\!\alpha c}{f^{c}}_{\!\!ab}+
{F^{\gamma}}_{\!\!\alpha\beta}{K^{\gamma}}_{\!\!ab})(-1)^{\varepsilon_{\alpha}\varepsilon_b}+
{K^{\gamma}}_{\!\!ab,j}R^j_{\alpha}
(-1)^{\varepsilon_{\alpha}\varepsilon_a}+\\
\nonumber
&&+({U^{\gamma}}_{\!\!\alpha a,j}T^j_b-
{U^{\gamma}}_{\!\!\alpha b,j}T^j_a(-1)^{\varepsilon_a\varepsilon_{\beta}})
(-1)^{\varepsilon_{\alpha}\varepsilon_b}-({U^{\gamma}}_{\!\!a \beta }{U^{\beta}}_{\!\!b \alpha }
-{U^{\gamma}}_{\!\!b\beta }{U^{\beta}}_{\!\!a\alpha }
(-1)^{\varepsilon_a\varepsilon_b})(-1)^{\varepsilon_{\alpha}\varepsilon_a}\big)+\\
\label{y1}
&&+\big(\lambda^{ij}_{\;\;ab,k}R^k_{\alpha}(-1)^{\varepsilon_{\alpha}(\varepsilon_a+\varepsilon_j)}
-R^i_{\alpha,k}\lambda^{kj}_{\;\;ab}(-1)^{\varepsilon_{\alpha}\varepsilon_b}\big)S_{0,j}+
\lambda^{ij}_{\;\;ab}S_{0,jk}R^k_{\alpha}(-1)^{\varepsilon_{\alpha}\varepsilon_a}=0.
\eeq
For closed global transformations ($\lambda^{ij}_{\;\;ab}=0$) it follows from (\ref{y1})
that the structure coefficients
${f^{c}}_{\!\!a b}$ are gauge invariant,
\beq
\label{g7g}
{f^{c}}_{\!\!a b,i}R^i_{\alpha}=0,
\eeq
and the following relations take place
\beq
\nonumber
&&-\big({U^{\gamma}}_{\!\!a \beta }{U^{\beta}}_{\!\!b \alpha }
-{U^{\gamma}}_{\!\!b\beta }{U^{\beta}}_{\!\!a\alpha }
(-1)^{\varepsilon_a\varepsilon_b}\big)
(-1)^{\varepsilon_{\alpha}(\varepsilon_a+\varepsilon_b)}+
{K^{\gamma}}_{\!\!ab,j}R^j_{\alpha}
(-1)^{\varepsilon_{\alpha}(\varepsilon_a+\varepsilon_b)}+\\
&&+
\big({U^{\gamma}}_{\!\!\alpha a,j}T^j_b-
{U^{\gamma}}_{\!\!\alpha b,j}T^j_a(-1)^{\varepsilon_a\varepsilon_b}\big)+
{U^{\gamma}}_{\!\!\alpha c}{f^{c}}_{\!\!ab}+
{F^{\gamma}}_{\!\!\alpha\beta}{K^{\gamma}}_{\!\!ab}=0.
\label{g7h}
\eeq
The Jacobi identity for operators ${\hat R}_{\alpha}, {\hat R}_{\beta}, {\hat T}_{a}$
lead to the relations
\beq
\nonumber
&&{F^{\sigma}}_{\!\!\alpha\beta,j}T^j_a-
{U^{\sigma}}_{\!\!a\gamma}{F^{\gamma}}_{\!\!\alpha\beta}
(-1)^{\varepsilon_a(\varepsilon_{\alpha}+\varepsilon_{\beta})}+\\
\nonumber
&&+
{F^{\sigma}}_{\!\!\alpha\gamma}{U^{\gamma}}_{\!\!\beta a}-{F^{\sigma}}_{\!\!\beta\gamma}
{U^{\gamma}}_{\!\!\alpha a}(-1)^{\varepsilon_{\alpha}\varepsilon_{\beta}}+\\
&&+{U^{\sigma}}_{\!\!\beta a,j}R^j_{\alpha}
(-1)^{\varepsilon_{\alpha}(\varepsilon_{\beta}+\varepsilon_{a})}-
{U^{\sigma}}_{\!\!\alpha a,j}R^j_{\beta}
(-1)^{\varepsilon_{a}\varepsilon_{\beta}}=0.
\label{g7h1}
\eeq
The relations(\ref{g1}), (\ref{g1c}) - (\ref{g2a}), (\ref{g7}) - (\ref{g7d0}), (\ref{y1})
and (\ref{g7h1}) describe structure and properties of symmetry algebra of the gauge system
under consideration.

The operators ${\hat R}$ and ${\hat T}_a$ do not commute,
\beq
\nonumber
[{\hat R},{\hat T}_a]&=&-{\hat R}_{\beta}
{U^{\beta}}_{\!\!\alpha a}C^{\alpha}(-1)^{\varepsilon_a}+\\
&&+ \frac{\overleftarrow{\delta}}{\delta
C^{\alpha}}\frac{1}{2}(-1)^{\varepsilon_{\beta}}
{F^{\alpha}}_{\!\!\beta\gamma,j}T^j_{a}C^{\gamma}\;\!C^{\beta}
(-1)^{\varepsilon_a(\varepsilon_{\beta}+\varepsilon_{\gamma})}.
\label{g7f1}
\eeq
From (\ref{g7f1}) we obtain  the important relations
\beq
\label{g7f2}
\Psi[{\hat R},{\hat T}_a ]=-\Psi{\hat R}_{\beta}
{U^{\beta}}_{\!\!\alpha a}\;\!C^{\alpha}(-1)^{\varepsilon_a}.
\eeq
The right-hand side in (\ref{g7f2}) is nothing but the gauge
transformations of $\Psi$ with gauge
parameters $\Lambda^{\beta}(\phi)={U^{\beta}}_{\!\!\alpha
a}(\varphi)C^{\alpha}(-1)^{\varepsilon_a}\omega^a$.

The quantum action (\ref{g4}) is not invariant under the global
transformations (\ref{g7}). Using (\ref{g7f2}), the variation of $S=S(\phi)$
can be presented in the form
\beq
\label{g9a}
\delta_T S= S{\hat
T}_a\omega^a=\big(\Psi{\hat T}_a\omega^a\big){\hat R}+
\big(\Psi{\hat R}_{\beta}\big){U^{\beta}}_{\!\!\alpha a}
\omega^aC^{\alpha}.
\eeq
The first term in the right-hand side of
(\ref{g9a}) describes the variation of gauge fixing functional $\Psi$ under
global transformation while the second summand is the gauge
transformation of $\Psi=\Psi(\phi)$ with local parameters
$\Lambda^{\alpha}=\Lambda^{\alpha}(\phi)$,
\beq
\label{g10}
\Lambda^{\alpha}(\phi)= {U^{\alpha}}_{\!\!\beta
a}(\varphi)\omega^aC^{\beta}.
\eeq
Second term in (\ref{g9a})
can be presented in the form
\beq
\label{g11} \big(\Psi{\hat
R}_{\beta}\big) {U^{\beta}}_{\!\!\alpha a}
\omega^aC^{\alpha}=S_{,\alpha} \Lambda^{\alpha},\quad
S_{,\alpha}=S\frac{\overleftarrow{\delta}}{\delta C^{\alpha}}.
\eeq
It will allow us in the next Section to analyze the (in)dependence of the effective action
on the global symmetry transformations in the theory under consideration.

\section{Structure of global symmetries on quantum level}

In this Section we consider properties  of global symmetry within
the BRST quantization  taking into account that the status of the gauge symmetry
is well-known. In particular, the vacuum functional, Z,
\beq
\label{g9}
Z=Z_{\chi}=\int D\phi \exp\Big\{\frac{i}{\hbar} S(\phi)\Big\}
\eeq
does not depend on the choice of admissible gauge fixing functions
$\chi_{\alpha}$  thanks to the BRST symmetry of $S(\phi)$ (\ref{g5}),
\beq
\label{g10}
\delta_{\chi} Z=0.
\eeq
In deriving this result the
following conditions
\beq
\label{g11}
(-1)^{\varepsilon_{\beta}}{F^{\beta}}_{\!\!\beta\alpha}(\varphi)=0,\quad
(-1)^{\varepsilon_i}\frac{\overrightarrow{\delta}}{\delta
\varphi^i}R^i_{\alpha}(\varphi)=0 \eeq
are used.

Let $Z_T$ be the vacuum functional for the theory with action
$S(\phi)+\delta_T S(\phi)$
\beq
\label{g12}
Z_T=\int D\phi
\exp\Big\{\frac{i}{\hbar} \big[S(\phi)+\delta_T S(\phi)\big]\Big\}.
\eeq
Making use in the functional integral (\ref{g12}) the change of variables in
the form of BRST
transformations (\ref{g5}) but with replacement $\lambda \rightarrow  \Lambda (\phi)$
 where
\beq
\label{g13}
\Lambda (\phi)=\frac{i}{\hbar}{\bar C}^{\alpha}
\chi_{\alpha,i}(\varphi)T^i_a(\varphi)\omega^a=
\frac{i}{\hbar}\Psi(\phi){\hat T}(\varphi),
\eeq
and taking into account the triviality of Jacobian of such change, we arrive at the relation
\beq
\label{g13a}
Z_T=\int D\phi
\exp\Big\{\frac{i}{\hbar} \big[S(\phi)+ S_{,\alpha}(\phi)\Lambda^{\alpha}(\phi)\big]\Big\}.
\eeq
Then performing the change of  variables  $C^{\alpha}$ in the form
\beq
\label{g13b}
C^{\alpha} \;\rightarrow  \; C^{\alpha} -
{U^{\alpha}}_{\!\!\beta a}(\varphi)\omega^aC^{\beta}
\eeq
with the Jacobian equal to the unit and assuming the fulfilment of the conditions
\beq
\label{g13c}
(-1)^{\varepsilon_{\alpha}}{U^{\alpha}}_{\!\!\alpha a}(\varphi)=0,
\eeq
we have the statement
\beq
\label{g14}
 Z_T=Z , %\quad {\rm or}\quad \delta_T Z=0,
\eeq
i. e. the vacuum functional is invariant under the  classical global transformations.

The generating functional of the Green functions, $Z(J)$, and the connected
Green functions, $W(J)$, is represented
by the  functional integral
\beq
\label{g15}
Z(J)=\int D \phi\exp\Big\{\frac{i}{\hbar}\big[S(\phi)
+J_A\phi^A\big]\Big\}=\exp\Big\{\frac{i}{\hbar}W(J) \Big\}.
\eeq
As a main consequence of the BRST symmetry of $S(\phi)$, there exits the Ward identity
for $Z(J)$ and $W(J)$ in the form
\beq
\label{g15a}
J_AR^A\Big(\frac{\hbar}{i}\frac{\delta}{\delta J}\Big)Z(J)=0,\quad
J_AR^A\Big(\frac{\delta W}{\delta J}+\frac{\hbar}{i}\frac{\delta}{\delta J}\Big)\cdot 1=0.
\eeq
The generating functional of vertex functions (effective action), $\Gamma=\Gamma(\Phi)$,
is defined standardly through the Legendre transformation of $W(J)$,
\beq
\label{g22}
\Gamma(\Phi)= W(J)-J_A\Phi^A,\quad
\Phi^A=\frac{\delta{W}(J)}{\delta J_A},
\eeq
so that
\beq
\label{g23}
\Gamma(\Phi)\frac{\overleftarrow{\delta}}{\delta\Phi^A} =-J_A.
\eeq
The Ward identity (\ref{g15a}) rewrites for $\Gamma(\Phi)$ in the form
\beq
\label{g15b}
\Gamma(\Phi)\frac{\overleftarrow{\delta}}{\delta\Phi^A}{\bar R}^A(\Phi)=0,
\eeq
where
\beq
\label{g15c}
{\bar R}^A(\Phi)=R^A({\hat \Phi})\cdot 1,
\eeq
and
\beq
\label{g26}
{\hat \Phi}^A=\Phi^A+i\hbar
(\Gamma^{'' -1})^{AB}(\Phi)\frac{\overrightarrow{\delta}}{\delta\Phi^B}.
\eeq
In (\ref{g26}) the matrix $(\Gamma^{'' -1})^{AB}(\Phi)$ is inverse to
\beq
\label{g27}
(\Gamma^{''})_{AB}(\Phi)=
\frac{\overrightarrow{\delta}}{\delta\Phi^A}\Big(\Gamma(\Phi)
\frac{\overleftarrow{\delta}}{\delta\Phi^B}\Big),\quad
(\Gamma^{'' -1})^{AC}(\Gamma^{''})_{CB}=\delta^A_B .
\eeq
The Ward identity (\ref{g15b}) can be interpreted as the invariance of
effective action $\Gamma(\Phi)$ under the {\it quantum} BRST transformations of $\Phi^A$
with generators ${\bar R}^A(\Phi)$.

In the functional integral (\ref{g15}) we make the change of integration
variables $\varphi^i$
in the form of global transformations (\ref{g7}). Then, using the conditions
\beq
\label{g16}
(-1)^{\varepsilon_i}\frac{\overrightarrow{\delta}}{\delta
\varphi^i}T^i_{a}(\varphi)=0,
\eeq
we obtain
\beq
\label{g17}
Z(J)=\int D \phi\exp\Big\{\frac{i}{\hbar}\big[S(\phi)+\delta_T S(\phi)
+J_A\phi^A+j_iT^i_a(\varphi)\omega^a\big]\Big\},
\eeq
where $\delta_T S(\phi)$ is defined in (\ref{g9a}) and $j_i$
are external sources to $\varphi^i$. The conditions (\ref{g16}) lead to that the Jacobian of change of variables
in the functional integral (\ref{g17}) is equal to unit.
By performing the change of variables $\phi^A$ in the form
\beq
\label{g18}
\phi^A \;\rightarrow \;\phi^A+R^A(\phi)\Lambda(\phi)
\eeq
with $\Lambda(\phi)$ given in (\ref{g13}), and
then additionally the transformations (\ref{g13b}),
we find in the first order in $\omega^a$
\beq
\label{g19}
\int D \phi\Big(j_iT^i_a(\varphi)\!+\!J_AR^A(\phi)\Lambda_a(\phi)\!+\!
J_{\alpha_{(C)}}{U^{\alpha}}_{\!\!a \beta}(\varphi)C^{\beta}(-1)^{\varepsilon_a}\Big)
\exp\Big\{\frac{i}{\hbar}\big[S(\phi)\!+\!J_A\phi^A\big]\Big\}=0,
\eeq
where we took into account that the Jacobian of change of variables is trivial and $J_{\alpha_{(C)}}$ are sources
to fields $C^{\alpha}$.
The equation (\ref{g19}) rewrites
\beq
\label{g20}
&&\!\!\!\!\!\!\!\!\!\!\!\Big[j_iT^i_a\Big(\frac{\hbar}{i}\frac{\delta}{\delta j}\Big)\!+\!
J_AR^A\Big(\frac{\hbar}{i}\frac{\delta}{\delta J}\Big)
\Lambda_a\Big(\frac{\hbar}{i}\frac{\delta}{\delta J}\Big)\!+
\!\frac{\hbar}{i}J_{\alpha_{(C)}}{U^{\alpha}}_{\!\!a \beta}
\Big(\frac{\hbar}{i}\frac{\delta}{\delta j}\Big)
\frac{\delta}{\delta J_{\beta_{(C)}}}
(-1)^{\varepsilon_a}\!\Big]Z(J)\!=\!0,\\
\nonumber
&& a=1,2,...,m.
\eeq
In terms of the functional $W=W(J)$  the relations (\ref{g20}) read
\beq
\nonumber
&&j_iT^i_a\Big(\frac{\delta W}{\delta j}\!+\!
\frac{\hbar}{i}\frac{\delta}{\delta j}\Big)\!+\!
J_AR^A\Big(\frac{\delta W}{\delta J}\!+\!\frac{\hbar}{i}\frac{\delta}{\delta J}\Big)
\Lambda_a\Big(\frac{\delta W}{\delta J}\!+\!
\frac{\hbar}{i}\frac{\delta}{\delta J}\Big)\!+\!\\
&&+
J_{\alpha_{(C)}}{U^{\alpha}}_{\!\!a \beta}
\Big(\frac{\delta W}{\delta j}\!+\!\frac{\hbar}{i}\frac{\delta}{\delta j}\Big)
\frac{\delta W}{\delta J_{\beta_{(C)}}}
(-1)^{\varepsilon_a}\!=\!0 , \quad  a=1,2,...,m..
\label{g21}
\eeq

Then in terms of $\Gamma(\Phi)$ the relations (\ref{g21}) can be presented in the form
\beq
\label{g24}
\Gamma(\Phi)\Big(\frac{\overleftarrow{\delta}}{\delta\Phi^i}T^i_a(\Phi)+
\frac{\overleftarrow{\delta}}{\delta\Phi^A}M^A_a(\Phi)+
\frac{\overleftarrow{\delta}}{\delta\Phi^{\alpha_{(C)}}}{U^{\alpha}}_{\!\!a}(\Phi)\Big)=0,
\eeq
where the notations
\beq
\nonumber
&&T^i_a(\Phi)=T^i_a(\varphi)|_{\varphi^i\rightarrow {\hat \Phi}^i}\!\!\cdot 1,\quad
M^A_a(\Phi)=R^A({\hat \Phi})\Lambda_a({\hat \Phi})\cdot 1,\;\\
\label{g25}
&&{U^{\alpha}}_{\!\!a}(\Phi)={U^{\alpha}}_{\!\!a \beta}(\varphi)|_{\varphi^i\rightarrow
{\hat \Phi}^i}\!\Phi^{\beta_{(C)}}(-1)^{\varepsilon_a} ,\\
\nonumber
&&
J_A=\big(j_i, J_{\alpha_{(B)}},J_{\alpha_{(C)}}, J_{\alpha_{({\bar C})}}\big),\quad
\Phi^A=\big(\Phi^i, \Phi^{\alpha_{(B)}},\Phi^{\alpha_{(C)}}, \Phi^{\alpha_{({\bar C})}}\big),
\eeq
were used.

The relations (\ref{g24}) mean that  the effective action is invariant under the
{\it quantum} global transformation,
\beq
\label{g28}
\Gamma(\Phi)\frac{\overleftarrow{\delta}}{\delta\Phi^A}{\bar T}^A_a(\Phi)=0
\eeq
with the deformed generators
\beq
\label{g29}
{\bar T}^A_a(\Phi) =\big(T^i_a(\Phi)+M^i_a(\Phi),\;0\;,
M^{\alpha_{(C)}}_a(\Phi)+U^{\alpha}_a(\Phi) , M^{\alpha_{({\bar C})}}_a(\Phi)\big).
\eeq
The proof of invariance (\ref{g28}) is based on using the change of variables (\ref{g13})
which are not analytical in loop expansion parameter $\hbar$. Therefore unlike the study of
gauge dependence of effective action in the framework of BRST-BV formalism, the derivation of
(\ref{g29}) is not related to change of gauge functions. In the next Section we
will however proof the correctness of (\ref{g28}) in loop expansion procedure.

\section{Deformation of the global transformations in one-loop approximation}

As we pointed out at the end of Section 4, the change of variables (\ref{g13})
is non-analytical in $\hbar$
and the use of loop expansion looks doubtful. However, we will show that
the one-loop approximation still works.

Consider  the relations (\ref{g28}) in loop approximation. For the effective action we have
\beq
\label{g30}
\Gamma(\Phi)=S(\Phi) +\hbar \Gamma_1(\Phi)+O(\hbar^2).
\eeq
The generators ${\bar T}^A_a(\Phi)$ are written in the same approximation as follows
\beq
\label{g31}
{\bar T}^A_a(\Phi)=\frac{1}{\hbar} R^A(\Phi) {\tilde \Lambda}(\Phi) +
{\bar T}^A_{(0)a}(\Phi)+\hbar {\bar T}^A_{(1)a}(\Phi)+O(\hbar^2),
\quad {\tilde \Lambda}(\Phi) =
\hbar \Lambda(\Phi).
\eeq
Due to the invariance of quantum action $S(\Phi)$ (\ref{g4}) under the BRST transformations (\ref{g5}),
the first item in the right-hand side of (\ref{g31}), which is non-analytical in $\hbar$,
does not contribute in the relations (\ref{g28}).
As a result
in zero loop approximations these relation take the form
\beq
\label{g32}
\frac{\delta S(\Phi)}{\delta\Phi^A}{\bar
T}^A_{(0)a}(\Phi)+
\frac{\delta\Gamma_1(\Phi)}{\delta\Phi^A}R^A(\Phi) {\tilde
\Lambda}(\Phi)=0. \eeq
In the next order we get
\beq \label{g32a}
S(\Phi)\frac{\overleftarrow{\delta}}{\delta\Phi^A}{\bar
T}^A_{(1)a}(\Phi)+
\Gamma_1(\Phi)\frac{\overleftarrow{\delta}}{\delta\Phi^A}{\bar
T}^A_{(0)a}(\Phi)=0.
\eeq
Now let us take into account the Ward
identity for $\Gamma(\Phi)$ (\ref{g15b}), then in the first order in
$\hbar$ we have \beq \label{g32c}
\Gamma_1(\Phi)\frac{\overleftarrow{\delta}}{\delta
\Phi^A}R^A(\Phi)=0. \eeq With this result the relations (\ref{g32})
rewrites \beq \label{g32d}
S(\Phi)\frac{\overleftarrow{\delta}}{\delta\Phi^A}{\bar
T}^A_{(0)a}(\Phi)=0.
\eeq
These relations demonstrate that the quantum action $S(\Phi)$ is one-loop invariant under the
deformed global symmetry transformations.
 In particular, the
generators of this global symmetry in the sector of fields $\Phi^i$
have the form
\beq
\label{g33} {\bar
T}^i_{(0)a}(\Phi)={\bar
T}^i_{(0)a}(\phi)|_{\phi \rightarrow \Phi},\quad {\bar
T}^i_{(0)a}(\phi)=
T^i_{a}(\varphi)-R^i_{\alpha}(\varphi)
\Sigma^{\alpha}_{(0)a}(\phi)-R^i_{\alpha,j}(\varphi)\Theta^{j\;\!\alpha}_{(0)\;\!a}(\phi),
\eeq
where the following notations
\beq
\nonumber
&&\Sigma^{\alpha}_{(0)a}(\phi)=
\left(C^{\alpha} (S^{'' -1})^{\beta_{({\bar C})}j }(\phi)+(S^{''
-1})^{\alpha_{(C)}j}(\phi){\bar C}^{\beta}
(-1)^{\varepsilon_j(\varepsilon_{\beta}+1)}\right)N_{\beta a,j}(\varphi)+\\
\nonumber
&&
\qquad\qquad\;\; +(S^{'' -1})^{\alpha_{(C)}\beta_{({\bar C})}}(\phi)
\chi_{\beta,kj}(\varphi)T^j_a(\varphi)+\\
\label{g34}
&&
\qquad\qquad\;\; +C^{\alpha}{\bar C}^{\beta}\chi_{\beta,j\;\!k}(\varphi)
(S^{'' -1})^{kl }(\phi)T^j_{a,\;\!l}(\varphi)
(-1)^{\varepsilon_l(\varepsilon_j+\varepsilon_a+1)},\\
\nonumber
&&
\Theta^{j\;\!\alpha}_{(0)\;\!a}(\phi)=
\left((S^{'' -1})^{j \;\!\alpha_{(C)}}(\phi){\bar C}^{\beta}+
(S^{'' -1})^{j\;\! \beta_{( {\bar C})}}(\phi)C^{\alpha}
(-1)^{(\varepsilon_{\alpha}+1)(\varepsilon_{\beta}+1)}\right)N_{\beta a}(\varphi)+\\
\label{g35}
&&\qquad\qquad\;\; +
(S^{'' -1})^{j\;l}(\phi)C^{\alpha}
{\bar C}^{\beta}N_{\beta a,l}(\varphi)
(-1)^{\varepsilon_l(\varepsilon_{\alpha}+\varepsilon_{\beta})},\\
\label{g36}
&&N_{\beta a}(\varphi)=\chi_{\beta,k}(\varphi)
T^k_a(\varphi).
\eeq
are introduced.
The deformation of global generators $T^i_{a}(\varphi)$, defined in initial configuration space
$\{\varphi\}$, looks very non-trivial already in the one-loop approximation. Such a deformation
is done in the full configuration space of the gauge theory (\ref{g3})
with the help of generators $R^i_{\alpha}(\varphi)$ of gauge symmetry of
initial action (\ref{g1})
and its derivatives $R^i_{\alpha,j}(\varphi)$. The relations (\ref{g1}), (\ref{g7}) and
(\ref{g33}) lead to
\beq
\label{g37}
S_{0,i}(\varphi){\bar
T}^i_{(0)a}(\phi)=
-S_{0,i}(\varphi)R^i_{\alpha,j}(\varphi)\Theta^{j\;\!\alpha}_{(0)\;\!a}(\phi)\neq 0.
\eeq
It means non invariance of an initial action
$S_0(\varphi)$ under the deformed global transformations.

\section{Summary}

In the present paper we have studied the general problem of global
(rigid) symmetries in quantum general gauge theory with closed
irreducible gauge algebra. Using the BRST-BV quantization procedure
\cite{BV,BV1} we have constructed a deformation of global symmetry
generators so that the full quantum action of the initial gauge
theory and hence the effective action are invariant under such
deformed global transformations. This statement is valid for any
bosonic or fermionic gauge theory with an irreducible and closed
algebra of gauge transformations and with an  arbitrary (even open)
non-anomalous algebra of global symmetry. Form of the deformed
global symmetry generators in one-loop approximation is calculated
in the explicit form. Note that from algebraic point of view the
deformation of global symmetry generators, studied in the present
paper, is similar to the known phenomena of the deformation of gauge
generators under renormalization procedure (see e.g.
\cite{LT3,LT1,VLT}), although the mechanisms of deformation differ.

In general the BRST-BV technique involves introduction for every
field of the full configuration space the corresponding antifield
with opposite Grassmann parity. It allows to study the  many
properties of the gauge theory on quantum level in general terms.
However, to simplify the consideration in this paper we derived all
the results in the configuration space of fields only (\ref{g3}).
Generalization of the results obtained for extended configuration
space of fields and antifields can be done in the same method.

Problem of global symmetries in quantum general gauge theories is
discussed in earlier papers \cite{dWF,BPT,vanH,BHW} however the
accents were aimed on the other aspects. First, in \cite{dWF} it was
shown the possibility to construct the Lagrangian models possessing
the gauge and supersymmetry invariance under the assumption that the
global supersymmetry transformations obey the closed algebra. In our
paper we work with an arbitrary (including open) global
(super)symmetry of a given gauge theory. Also we suppose the absence
of anomalies in combined gauge and global algebra of generators but
in principle it is possible to study the problem of coexistence of
gauge and global symmetries in presence of anomalies using the
approach of work \cite{BPT}. Of course the implementation of this
program requires a separate independent study. Second, in paper
\cite{vanH} it was shown that the attempts to construct an action
invariant under BRST- and closed rigid symmetries lead to the
breakdown of non-degeneracy of the full quantum action. In our paper
we proved that this problem is resolved on the base of quantum
deformation of global generators. Third, in paper \cite{BHW} it was
shown that the global symmetries of an initial gauge action in the
field-antifield formalism can be extended to include the all fields
and the all antifields. After that the authors of \cite{BHW}
introduced a constant ghost for each global symmetry and modified
the master-equation to incorporate the global symmetries into its
solutions but they did not discuss the deformations of the global
symmetry generators. It is clear that this approach differs from our
consideration since from the beginning we work only in the
configuration field space and derive the above deformations.

The general enough approach to the problem of global symmetries in
the quantum gauge theory was proposed in the work \cite{K1} under
assumptions that the initial classical theory consists of only
bosonic fields and the algebra of the classical global
transformations is closed. In our paper we have considered maximally
general case of the theory with bosonic and fermionic fields and
open algebra of the global (super)symmetries. However, what is very
important, we work with more general effective action then in
\cite{K1}. In our case, the sources included not only to initial
fields but also to ghost and auxiliary fields. This allowed us to
describe the deformation of the generators completely in the
algebraic terms.

The global (super)symmetric  and gauge transformations have
been studied in the framework of singular gauge fixing procedure.
In practice it is more convenient to use a non-singular gauges.
All our basic statements still will be valid in this case as well. We should only  modify
the quantum action (\ref{g4}) in the form
\beq
\label{d1}
S(\phi)=S_0(\varphi)+\bar{C}^{\alpha}\chi_{\alpha
,i}(\varphi)R^i_{\beta}(\varphi)C^{\beta} +
\chi_{\alpha}(\varphi)B^{\alpha}+\frac{\xi}{2}B^{\alpha}B^{\alpha},
\eeq
where $\xi$ is a gauge parameter. Due to the property of BRST transformations
(\ref{g6}) $B^{\alpha}{\hat R}=0$, the action (\ref{d1}) remains invariant under the BRST
transformations, $S{\hat R}=0$. Thus, from a principled point of view the results obtained will be same.

\section*{Acknowledgments}
\noindent The authors thank I.V. Tyutin for useful discussions and
S.M. Kuzenko for correspondence. The research was supported in parts
by Russian Ministry of Education and Science, project No.
3.1386.2017. The authors are also grateful to RFBR grant,
 project No. 18-02-00153 for partial support.

\end{document}